\documentclass[epj]{svjourliza}
\usepackage{graphics,epsfig}
\usepackage{amsmath,amsfonts}
\usepackage{latexsym}
\newcommand{\bra}[1]{\langle #1|}
\newcommand{\ket}[1]{|#1\rangle}

\newcommand{\im}{\textrm{Im}\ }
\begin{document}
\title{Superfrustration of charge degrees of freedom\thanks{Contribution to the proceedings of the XXIII
IUPAP International Conference on Statistical Physics in Genova, Italy.}}
\author{
Liza Huijse
\and
Kareljan Schoutens
}                     
%
\institute{
Institute for Theoretical Physics, University of Amsterdam,
Valckenierstraat 65, 1018 XE Amsterdam, the Netherlands
}
\date{Date: September 5, 2007}
%
\abstract{
We review recent results, obtained with P.~Fendley, on 
frustration of quantum charges in lattice models for
itinerant fermions with strong repulsive interactions.
A judicious tuning of kinetic and interaction terms leads
to models possessing supersymmetry. In such models frustration
takes the form of what we call superfrustration: 
an extensive degeneracy of supersymmetric ground states.
We present a gallery of examples of superfrustration on
a variety of 2D lattices.
\PACS{
      {05.30.-d}{Quantum statistical mechanics}   \and
      {11.30.Pb}{Supersymmetry} \and
      {71.27.+a}{Strongly correlated electron systems; heavy fermions}
     } 
} 
\maketitle
\section{Introduction}
\label{intro}
When charged particles, with repulsive interactions, are placed 
on a lattice one expects {\bf geometric frustration}:
depending on the lattice and the number of particles, there
can be many configurations that realize the lowest possible
interaction energy. Well-studied examples are charges on 
the triangular and checkerboard lattices (see \cite{early-frus} 
for some early references). In general, including kinetic terms 
for the quantum charges lifts the degeneracies. Depending on 
details, this may give rise to novel phases of quantum matter 
with remarkable physical properties \cite{frus}. The theoretical 
tools for studying these systems are limited: one typically relies 
on a strong coupling expansions and on numerics.

Recent work by P.~Fendley and one of the authors \cite{FS} has
uncovered models for strongly
interacting itinerant fermions which display a strong form
of quantum charge frustration, which we call {\bf superfrustration}.
These models (defined on 2D or 3D lattices) have a large, exact 
ground state degeneracy in the presence of kinetic terms.
Superfrustration thus arises due to a subtle interplay between 
kinetic terms and strong repulsive interactions.

The term `superfrustration' has its origin in a key property
used to identify the models and to study their properties, which 
is {\bf supersymmetry}. Quite remarkably, the notion of supersymmetry,
which was developed in the context of high energy physics, turns out 
to be a powerful tool in the analysis of strongly correlated itinerant 
fermions. This was first recognized in the context of 1D models
\cite{FSdB,FNS}. The extension to 2D and 3D then led to the discovery 
of the phenomenon of superfrustration.  

In this review, we shall first explain 
(Sect.~\ref{susy}) how supersymmetry 
is put to work in lattice models of correlated fermions. We introduce
some basic tools, such as the Witten index, make a connection
to cohomology theory and discuss a model on a 1D chain. We then move 
to models on 2D lattices (Sect.~\ref{beyond1D}), where 
we present a heuristic geometric intuition (3-rule) and discuss the 
methods employed in the analysis. In Sect.~\ref{beyond1Dex} we 
present a gallery 
of examples, each chosen such as to illuminate specific aspects and 
features. They firmly establish the notion of superfrustration in
its various guises. We close (Sect.~\ref{conclusion}) with some 
thoughts about the nature of the various ground states and quantum 
phases, in particular in relation to quantum criticality.

\section{Supersymmetry}
\label{susy}

\subsection{Basic algebra and Hamiltonian}
In quantum mechanics, supersymmetric theories are characterized by a positive 
definite energy spectrum and a twofold degeneracy of
each non-zero energy level. The two states with the same energy are called 
superpartners and are related by the nilpotent supercharge operator. 
Let us consider an $\mathcal{N}=2$ supersymmetric theory, defined by two
nilpotent supercharges $Q$ and $Q^{\dag}$ \cite{Wi},
\begin{eqnarray}\nonumber
Q^2=(Q^{\dag})^2=0
\end{eqnarray}
and the Hamiltonian given by 
\begin{eqnarray}\nonumber
H=\{Q^{\dag},Q\} .
\end{eqnarray}
From this definition it follows directly that $H$ is positive definite:
\begin{eqnarray*}
\bra{\psi}H\ket{\psi}
  &=& \bra{\psi}(Q^{\dag}Q+QQ^{\dag})\ket{\psi}
\nonumber \\[2mm]
  &=&|Q\ket{\psi}|^2+|Q^{\dag}\ket{\psi}|^2 \geq 0 \ .
\end{eqnarray*}
Furthermore, both $Q$ and $Q^{\dag}$ commute with the Hamiltonian, 
which gives rise to the twofold degeneracy in the energy spectrum. 
In other words, all eigenstates with an energy $E_s>0$ form doublet
representations of the supersymmetry algebra. A doublet consists of two 
states $\ket{s}, Q \ket{s}$, such that $Q^{\dag}\ket{s}=0$. Finally, 
all states with zero energy must be singlets: 
$Q \ket{g}=Q^{\dag}\ket{g}=0$ and conversely, all singlets must be zero 
energy states \cite{Wi}. In addition to supersymmetry our models also 
have a fermion-number symmetry generated by the operator $F$ with
\begin{eqnarray}\nonumber
[F,Q^{\dag}]=-Q^{\dag} \quad \textrm{and} \quad [F,Q]=Q.
\end{eqnarray}
Consequently, $F$ commutes with the Hamiltonian.

The supersymmetric theories that we discuss in this paper describe
fermionic particles. One might expect that, in order to be
supersymmetric, these theories would need bosonic particles as well
but this is not the case. The crux is that the quantum states in
these theories come in two types: bosonic states, having an even 
number $f$ of fermionic particles, and fermionic states with $f$ odd. 
From the commutators of $F$ with the supercharges, one finds that $Q$ 
and $Q^{\dag}$ change $f$ by plus or minus one one unit, so that the 
supercharges map bosonic states to fermionic states and vice versa.

We now make things concrete and define supersymmetric models for spin-less
fermions on a lattice or graph with $L$ sites in any dimension, following
\cite{FSdB}). The operator 
that creates a fermion on site $i$ is written as $c_i^{\dag}$ with
$\{c_i^{\dag},c_j\}=\delta_{ij}$. A simple choice for the first supercharge 
would be $Q=\sum_i c_i^{\dag}$, where the sum is over all lattice sites.
This leads to a trivial Hamiltonian: $H=L$, where $L$ is the number of 
lattice sites. To obtain a non-trivial Hamiltonian, we dress the fermion 
with a projection operator: 
$P_{<i>}=\prod_{j \textrm{ next to } i} (1-c_j^{\dag}c_j)$, which requires 
all sites adjacent to site $i$ to be empty. With $Q=\sum c_i^{\dag} P_{<i>}$ 
and $Q^{\dag}=\sum c_i P_{<i>}$, the Hamiltonian of these hard-core 
fermions reads
\begin{eqnarray}\label{Hsusygen}\nonumber
H=\{Q^{\dag},Q\}= 
\sum_i \sum_{j\textrm{ next to }i} P_{<i>} c_i^{\dag} c_j P_{<j>} 
       + \sum_i P_{<i>}.
\end{eqnarray}
The first term is just a nearest neighbor hopping term for hard-core fermions, 
the second term contains a next-nearest neighbor repulsion, a chemical 
potential and a constant. The details of the latter terms will depend on the 
lattice we choose. 

Note that all the parameters in the Hamiltonian (the hopping $t$, the nearest 
neighbor repulsion $V_1$, the next-nearest neighbor repulsion $V_2$ and the 
chemical potential $\mu$) are fixed by the choice of the supercharges and
the requirement of supersymmetry and eventually the lattice.

\subsection{Witten index}

We have already discussed how supersymmetry gives rise to certain properties
of the spectrum of the system, such as positivity of the energies and 
pairing of the excited states. An important issue is whether or not
supersymmetric ground states at zero energy occur. For this one considers  
the Witten index
\begin{equation}
W = \hbox{tr}\left[(-1)^F e^{-\beta H}\right] \ . \label{Windex}
\end{equation}
Remember that all excited states come in doublets with the same energy and 
differing in their fermion-number by one. This means that in the trace all 
contributions of excited states will cancel pairwise, and that the only
states contributing are the zero energy ground states. We can thus evaluate 
$W$ in the limit of $\beta \rightarrow 0$, where all states contribute 
$(-1)^F$. It also follows that $|W|$ is a lower bound to the number of 
zero energy ground states.

\subsection{Example: 6-site chain}
Let us consider as an example of all the above, the chain of six sites 
with periodic boundary conditions. The first thing we note is that
the Hamiltonian for an $L$-site chain with periodic boundary conditions 
can be rewritten in the following form:
\begin{eqnarray}
H &=& H_{\mathrm{kin}} + H_{\mathrm{pot}},
\label{ham1} 
\end{eqnarray}
where
\begin{eqnarray}\nonumber
H_{\mathrm{kin}}&=& \sum_{i=1}^{L} 
\left[ P_{i-1} \big(c_i^{\dag} c_{i+1} 
                    + c_{i+1}^{\dag} c_i \big)  P_{i+2} \right] , \\
H_{\mathrm{pot}}&=& \sum_{i=1}^{L} P_{i-1}P_{i+1} \nonumber\\
&=& \sum_{i=1}^{L} \left[ 1 -2 n_i + n_i n_{i+2}\right]  \nonumber\\
&=& \sum_{i=1}^{L} (n_i n_{i+2}) + L - 2 F .\nonumber
\end{eqnarray}
Here $P_i=1-n_i$, $n_i=c^{\dag}_i c_i$ is the usual number operator and
$F=\sum_i n_i$ is the total number of fermions. (We shall denote 
eigenvalues of this operator by $f$ and write the fermion density or
filling fraction as $\nu=f/L$.) The form of the hamiltonian makes clear 
that the hopping parameter $t$ is tuned to be equal to the next-nearest 
neighbor repulsion $V_2$, which is tuned to unity. The nearest neighbor 
repulsion $V_1$ is by definition infinite and the chemical potential $\mu$ 
is $2$. Finally, there is a constant contribution $L$ to the Hamiltonian. 
Note that the second term in the Hamiltonian $H_{\mathrm{pot}}$ suggests 
that the energy is minimized when the hard-core fermions are three sites 
apart.

\begin{figure}
\centering
\resizebox{0.8\columnwidth}{!}{%
\includegraphics{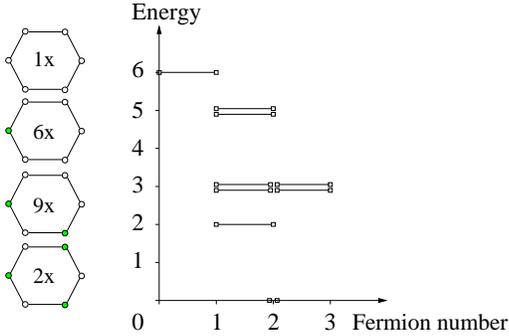}
}
\caption{Configurations and spectrum of the 6-site chain.}
\label{6site}       
\end{figure}

Let us consider the possible configurations of the 6-site chain. 
In addition to the empty state, there are six configurations with one 
fermion, nine with two fermions and two with three fermions 
(see Fig.~\ref{6site}). Because of the 
hard-core character of the fermions, half-filling is the maximal density. 
Clearly, the operator $Q$ gives zero on these maximally filled states. 
On the other hand, $Q^{\dag}$ acts non-trivially on these states, 
so two of the nine states with two fermions are superpartners of the 
maximally filled states. The empty state $\ket{0}$ has an energy $E=6$ 
and $Q^{\dag}\ket{0}=0$, whereas $Q \ket{0}= \sum_i c^{\dag}_i
\ket{0}$,  so $(\ket{0},Q \ket{0})$ make up a doublet. The other five 
states with one fermion are annihilated by $Q^{\dag}$ and $Q$ acts 
non-trivially on them, so they form supersymmetry doublets with five 
two-fermion-states. At this point, seven of the nine two-fermion-states 
are paired up in doublets, either with one- or three-fermion-states. 
The remaining two states cannot be part of a doublet, which implies that 
they must be singlet states and thus have zero energy. So we find that the 
6-site chain has a twofold degenerate zero energy ground state at 
filling $\nu=f/L = 1/3$. The full spectrum of the 6-site chain is shown
in Fig.~\ref{6site}. 

We observe that the ground state filling fraction of $1/3$ agrees with
the expectation that fermions tend to be three sites apart. This 
geometric rule suggests three possible ground states; in the full quantum
theory two are realized as zero-energy states. Note that the actual ground 
state wavefunctions are superpositions of many different configurations.
With a bit more work one can show that the ground states have eigenvalues 
$\exp(\pm \pi i/3)$ under translation by one site. 

Now let us compute the Witten index for this example. Remember that for a 
supersymmetric theory it simply reads
\begin{eqnarray}\nonumber
W= \textrm{tr}(-1)^F .
\end{eqnarray}
Note that we can take any basis of states we like to compute the trace. 
Above we have specified a basis by considering all the possible 
configurations of up to three fermions on the chain. It immediately gives
$W=1-6+9-2=2$ in agreement with the existence of the two ground states 
that we found. 

We close this section with two comments. First, we stress that the extremely
simple computation of $W$ alone guarantees the existence of at least two
ground states at zero energy. Similar results are easily established for 
much larger systems, where a direct evaluation of the ground state energies
is way out of reach, showing the power of supersymmetry. Second, we observe 
that here the Witten index is exactly equal to the number of ground states. 
We will encounter examples where ground states exist at more than one 
fermion number $f$, leading to cancellations in the Witten index so that
$|W|$ is strictly smaller than the number of ground states.

\subsection{Cohomology}
Supersymmetry supplies us with another tool, besides the Witten index, to 
study the ground states of the fermion models. This so-called cohomology 
method is more involved but it reveals more information about the ground 
state structure, in that it specifies the number of ground states for 
given fermion number $f$.

The key ingredient is the fact that ground states are singlets, they are 
annihilated both by $Q$ and $Q^{\dag}$. This means that a ground state 
$\ket{g}$ is in the kernel of $Q$:  $Q\ket{g}=0$. Such a ground state
is not in the image of $Q$, because if we could 
write $\ket{g}=Q\ket{f}$, then $(\ket{f}, \ket{g})$, would be a doublet. 
Equivalently, we can say that a 
ground state is closed but not exact. So the ground states span a subspace $H_Q$ of 
the Hilbert space $\mathcal{H}$ of states, such that $H_Q=\ker Q/\im Q$. This is precisely the definition 
of the cohomology of $Q$. So the ground states of a supersymmetric theory are in one-to-one correspondence 
with the cohomology of $Q$. Two states $\ket{s_1}$ and $\ket{s_2}$ are said to be in the same 
cohomology-class if $\ket{s_1}=\ket{s_2}+ Q \ket{s_3}$ for some state $\ket{s_3}$. Since a ground state is
annihilated by both $Q$ and $Q^{\dag}$, different (i.e. linearly independent) ground states must be in 
different cohomology-classes of $Q$. Finally, the number of independent ground states is precisely the dimension of the 
cohomology of $Q$ and the fermion-number of a ground state is the same as that of the corresponding 
cohomology-class.

There are several techniques to compute the cohomology, which we shall 
illustrate by working out examples in the following sections.   

\subsection{Example: 1D chains}
In previous work \cite{FS,FSdB,FNS,BDA} the supersymmetric model on the 
chain was studied extensively. We will summarize some of the results, but 
mostly use this case to illustrate the power of the tools we have
developed in the previous sections. Let us first compute the Witten index. 
In the example of the 6-site chain we saw that the Witten index can be 
computed by simply summing over all possible configurations with
the appropriate sign. However, because of the hard-core character of the 
fermions this is not a trivial problem for larger sizes. Here we shall 
exploit a much more elegant method, which will turn out very useful when 
we extend our model to more complex lattices. This method consists of the 
following steps: first divide the lattice into two sublattices $S_1$ and 
$S_2$. Then fix the configuration on $S_1$ and sum $(-1)^F$ for the
configurations on $S_2$. Finally, sum the results over the configurations 
of $S_1$. Of course the trick is to make a smart choice for the sublattices. 
For the periodic chain with $L=3j$ sites, we take $S_2$ to
be every third site. All the sites on $S_2$ are disconnected and thus 
every site can be either empty or occupied given that its neighboring sites 
on $S_1$ are empty. This means that the sum of $(-1)^F$ for the configurations 
on $S_2$ vanishes as soon as at least one site on $S_2$ can be both empty and
occupied. Consequently, the only non-zero contribution comes from the 
configurations such that at least one of the adjacent sites on $S_1$ is 
occupied. There are only two such configurations:
\begin{eqnarray}\label{configs}
\ket{\alpha} &\equiv& \dots \bullet \Box \circ \bullet \Box \circ \bullet \Box \circ \bullet \Box \circ \bullet 
\Box \circ \bullet \Box \circ \bullet \Box \circ \dots \nonumber\\
\ket{\gamma} &\equiv& \dots \circ \Box \bullet \circ \Box \bullet \circ \Box \bullet \circ \Box \bullet \circ
\Box \bullet \circ \Box \bullet \circ \Box \bullet \dots
\end{eqnarray}
where the square represents an empty site on $S_2$. The final step is to sum $(-1)^F$ for these two
configurations, and since both configurations have $f=L/3=j$, we find that the Witten index is $W=2(-1)^j$.
Note that this agrees with the result we obtained for the 6-site chain.

To find the exact number of ground states we compute the cohomology by using a \emph{spectral sequence}. 
A useful theorem is the 
`tic-tac-toe' lemma of \cite{BT}. This says that under certain conditions, the cohomology
$H_Q$ for $Q = Q_1 + Q_2$ is the same as the cohomology
of $Q_1$ acting on the cohomology of $Q_2$. In an equation, $H_Q = H_{Q_1}(H_{Q_2} ) \equiv H_{12}$, where 
$Q_1$ and $Q_2$ act on different sublattices $S_1$ and 
$S_2$. We find $H_{12}$ by first fixing the configuration on all sites of the sublattice $S_1$, and 
computing the cohomology $H_{Q_2}$. Then one computes the cohomology of $Q_1$, acting
not on the full space of states, but only on the classes
in $H_{Q_2}$. A sufficient condition for the lemma to hold is
that all non-trivial elements of $H_{12}$ have the same $f_2$ (the
fermion-number on $S_2$). For the periodic chain with $L=3j$ we choose the sublattice as before. Now 
consider a single site on $S_2$. If both of the adjacent $S_1$
sites are empty, $H_{Q_2}$ is trivial: $Q_2$ acting on the empty
site does not vanish, while the filled site is $Q_2$ acting on
the empty site. Thus $H_{Q_2}$ is non-trivial only when every
site on $S_2$ is forced to be empty by being adjacent to
an occupied site. The elements of $H_{Q_2}$ are just the two
states $\ket{\alpha}$ and $\ket{\gamma}$ pictured above in eq.~(\ref{configs}).
Both states $\ket{\alpha}$ and 
$\ket{\gamma}$ belong to $H_{12}$: they are closed because $Q_1 \ket{\alpha} =
Q_1 \ket{\gamma} = 0$, and not exact because there are no elements
of $H_{Q_2}$ with $f_1 = f -1$. By the tic-tac-toe lemma, there
must be precisely two different cohomology classes in $H_{Q}$,
and therefore exactly two ground states with $f = L/3$.
Applying the same arguments to the periodic chain with
$3f \pm 1$ sites and to the open chain yields in all cases
exactly one $E = 0$ ground state, except in open chains
with $3f + 1$ sites, where there are none \cite{FNS}.

The supersymmetric model on the chain can be solved exactly through 
a Bethe Ansatz \cite{FSdB}. In the continuum limit one can derive the 
thermodynamic Bethe ansatz equations. The model has the same
thermodynamic equations as the XXZ chain at $\Delta=-1/2$, so the two 
models coincide (the mapping can be found in \cite{FNS}). The
continuum limit of the XXZ chain is described by the massless
Thirring model \cite{hank}, or equivalently a free massless boson
$\Phi$ with action \cite{Friedan}
\begin{equation}\nonumber
S=\frac{g}{\pi}\int dx\,dt\ \left[(\partial_t \Phi)^2 -
(\partial_x\Phi)^2\right].
\end{equation}
The continuum limit of the $\Delta=-1/2$
model has $g=2/3$; this is the simplest field theory with ${\cal
N}=(2,2)$ superconformal symmetry \cite{Friedan}.  The $(2,2)$ means
that there are two left and two right-moving supersymmetries: in the
continuum limit the fermion decomposes into left- and right-moving
components over the Fermi sea. So the system is critical and in the 
continuum limit it is described by a superconformal field theory. 

A final note we would like to make here, is that the supersymmetric 
model on the chain recently emerged in a special limit of a large-$N$
supersymmetric matrix model \cite{VW}. This is yet another interesting 
connection, worthy of further investigation.

\section{Beyond 1D: heuristics and methodology}
\label{beyond1D}

We have seen that in the one dimensional case the Hamiltonian favors a 
configuration where the hard-core 
fermions sit three sites apart on average. This heuristic picture can be extended beyond 1D. For convenience
we restate the Hamiltonian in its general form
\begin{eqnarray}\nonumber
H= \sum_i \sum_{j\textrm{ next to }i} P_{<i>} c_i^{\dag} c_j P_{<j>} 
   + \sum_i P_{<i>}.
\end{eqnarray} 
The second term in the Hamiltonian gives a positive contribution to the 
energy for every site that has all neighboring sites empty, regardless
of whether the site itself is occupied or empty. For every hard-core 
fermion it will give a contribution of $+1$, since by definition it has 
its neighboring sites empty. So the contribution of this term is
minimized by blocking as many sites as possible with as few fermions 
as possible. Roughly speaking, this criterion leads to configurations
where all fermions are three sites apart. We call this the `3-rule'. 
In the following sections we shall see this 3-rule in action, and
demonstrate how it leads to superfrustration.

\vskip 3mm

In the previous section we have developed some tools to study our model 
on different lattices. The Witten index gives us a lower bound on the 
total number of supersymmetric ground states. In some cases one can
find a recursion relation or even a closed form for the Witten index 
as a function of the system size. The growth behavior of this function 
gives a lower bound to the growth behavior of the ground state entropy. 

If the Witten index grows exponentially with the system size, we have an 
extensive ground state entropy. In Sect.~\ref{martini} we present
an example where this entropy is known in closed form. 
Numerical studies of the Witten index on two-dimensional 
lattices \cite{vE} have revealed that for generic lattices the Witten 
index indeed 
grows exponentially with the system size (see Sect.~\ref{triangle}
for an example). The square lattice is an exception, since there the 
Witten index grows exponentially with the perimeter of the system. We 
shall touch upon some features of the square lattice in Sect.~\ref{square}.

Further insight in the ground state structure can be obtained from the
cohomology method. This gives the exact number of ground states and for
each of them the number of fermions.
A remarkable feature of our model that has been revealed by cohomology 
studies is that ground states typically occur at different fermion-numbers, 
or equivalently at different filling fractions. This implies that the
Witten index will typically underestimate the actual number of ground states. 
More remarkably, however, this also implies that one can add a particle 
to the system or extract a particle from the system
within a certain window of filling fractions without paying any energy. 

Finally, it has in many cases turned out to be possible to characterize 
supersymmetric ground states with the help of an `effective geometric 
picture'. In this, one establishes a (almost) 1-1 correspondence 
between quantum ground states and geometric configurations such as coverings
of the lattice by dimers or tiles of specific dimensions. Examples
are dimer coverings for the case of the martini lattice (Sect.~\ref{martini})
and rhombus tilings for the 2D square lattice (Sect.~\ref{square}).
The geometric picture is related to the heuristic 3-rule, but much more 
robust. It has been pioneered in \cite{FS} and in a remarkable
series of mathematical papers by J.~Jonsson \cite{Jo}. 

The fact that the ground states of strongly correlated quantum fermion 
models can be characterized by geometric means is quite deep and at
this time not fully understood. It suggests that further properties 
of these models (such as the excited state spectra) are tractable by 
similar means, which opens most interesting perspectives. 

It the next section we present a gallery of examples of 2D lattices,
and indicate what has been revealed about their ground state structure
using the various approaches described in this section. A more
systematic account is forthcoming \cite{FHHS}.
 
\section{Beyond 1D: examples}
\label{beyond1Dex}

\subsection{Martini lattice}\label{martini}
\begin{figure}
\centering
\resizebox{0.8\columnwidth}{!}{%
\includegraphics{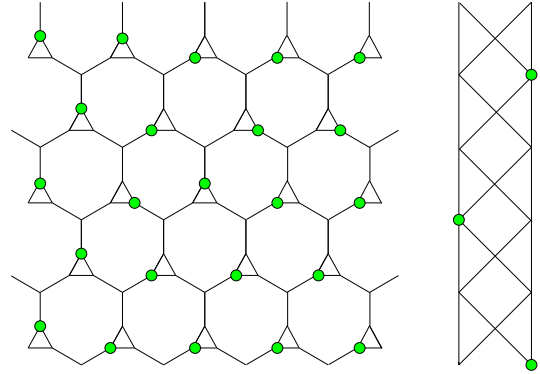}
}
\caption{Hard-core fermions on the martini lattice (left) and 
         on the kagome ladder (right).}
\label{martinilattice}       
\end{figure}

The martini lattice (see Fig. \ref{martinilattice}) is an example of a 
two dimensional lattice, where the cohomology can be computed relatively 
easily \cite{FS}. The method is strongly related to the one used to compute the
cohomology for the chain. The computation proves that the ground state 
entropy is an extensive quantity and we find a closed expression for the 
ground state entropy per site. We shall see that the 3-rule is not
violated in this case. This is related to the fact that the martini lattice, 
due to its structure, nicely accommodates the 3-rule. Lattices with a higher 
coordination number usually do not have this property and
consequently allow for a window of filling fraction for the supersymmetric 
ground states.

The martini lattice is formed by replacing every other site on a hexagonal 
lattice with a triangle. To find the ground states, take $S_1$ to be the 
sites on the triangles, and $S_2$ to be the remaining sites. As with the
chain, $H_{Q_2}$ vanishes unless every site in $S_2$ is adjacent to an 
occupied site on some triangle. The non-trivial elements of $H_{Q_2}$ 
therefore must have precisely one particle
per triangle, each adjacent to a different site on $S_2$. This is because 
a triangle can have at most one 
particle on it, and (with appropriate boundary conditions) there are
the same number of triangles as there are sites on $S_2$.
A typical element of $H_{Q_2}$ is shown in Fig. \ref{martinilattice}. One can
think of these as `dimer' configurations on the original
honeycomb lattice, where the dimer stretches from the
site replaced by the triangle to the adjacent non-triangle
site. Each close-packed hard-core dimer configuration is
in $H_{12}$, and by the tic-tac-toe lemma, it corresponds to
a ground state. The number of such ground states $e^{S_{\textrm{GS}}}$
is therefore equal to the number of such dimer coverings
of the honeycomb lattice, which for large $L$ is \cite{dimer}
\begin{equation}
\frac{S_{\textrm{GS}}}{L}=\frac{1}{\pi} \int_0^{\pi/3} d\theta 
     \ln[2 \cos \theta] = 0.16153 \dots \nonumber
\end{equation}
The frustration here clearly arises because there are many
ways of satisfying the 3-rule.

\subsection{Kagome ladder}
In this section we consider the kagome ladder 
(see Fig. \ref{martinilattice}) as an illustration of a case where we 
can compute the cohomology exactly, but where the 3-rule is not very 
helpful. We find a closed expression for the partition function
and a window of filling fraction for the supersymmetric ground states, 
which can both be interpreted in terms of tilings.

Computing the cohomology in this case is a bit more involved due to two 
things: First, in the previous
examples we could always choose the sublattice $S_2$ such that it 
consisted of disconnected sites, which by
themselves have zero cohomology. For the kagome ladder the convenient 
choice for the sublattice $S_2$ is less
trivial. The second complication arises because not all elements of 
the cohomology $H_{12}$ will have the same fermion-number, which was a 
sufficient condition for the tic-tac-toe lemma to hold.

\begin{figure}
\resizebox{0.8\columnwidth}{!}{%
\includegraphics{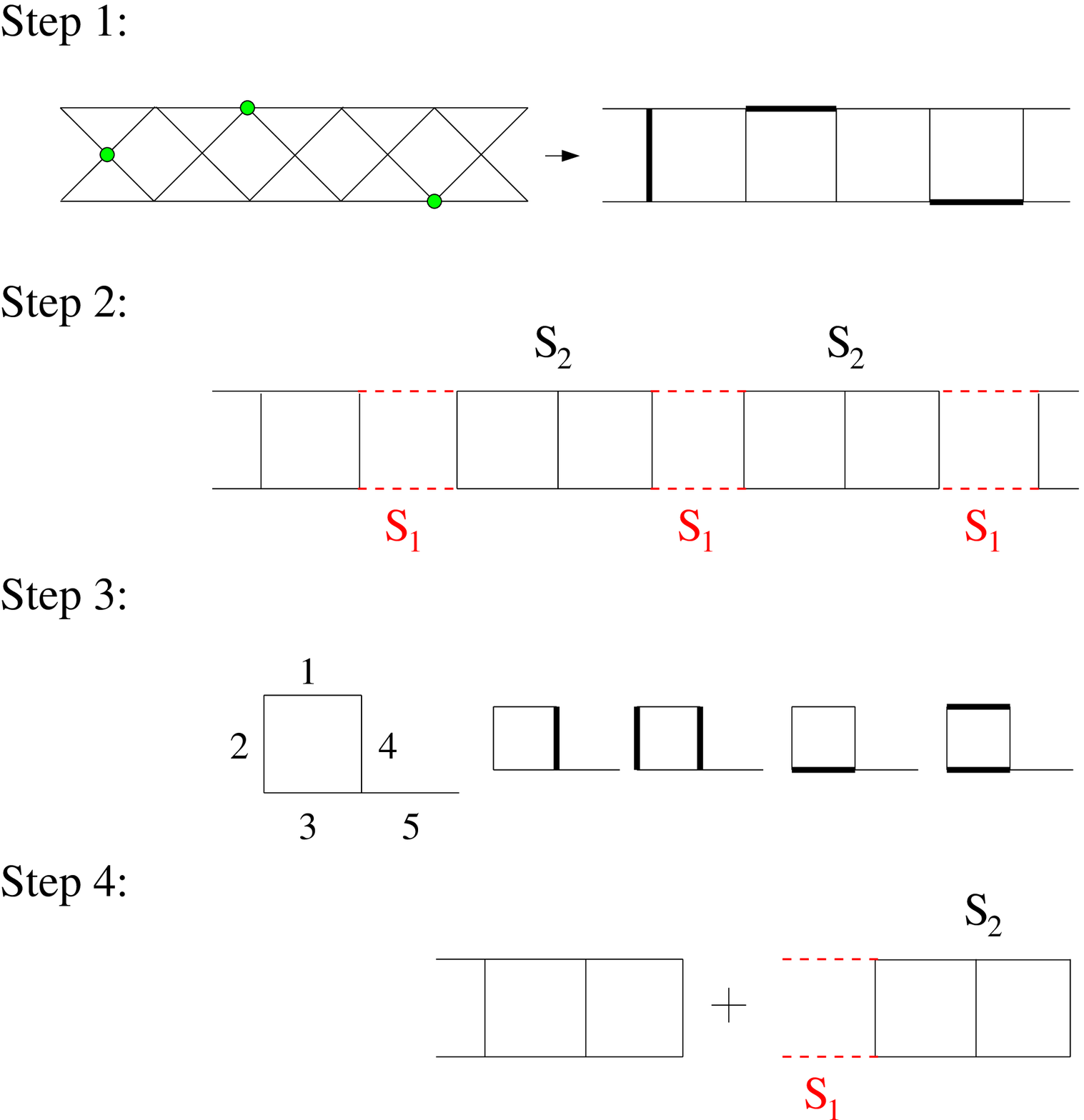}
}
\caption{Step-by-step computation of the cohomology of the kagome ladder.}
\label{kagomecohomology}       
\end{figure}

We compute the cohomology step by step. For each step there is a supporting 
picture in Fig.~\ref{kagomecohomology}. 
\textbf{Step 1} is to map the kagome ladder with hard-core fermions to a 
square ladder with hard-core dimers. This mapping is one-on-one. 
In \textbf{Step 2} we define the sublattices $S_1$ and $S_2$. 
\textbf{Step 3} is to note that the cohomology of one square plus one
additional edge vanishes. To do so, first note that if site 3 and 4 are 
both empty, we get zero cohomology due to site 5 which can now be both 
empty and occupied. The remaining four configurations are
easily shown to be either $Q$ of something (exact) or not in the kernel 
of $Q$ (not closed). \textbf{Step 4} is to build up the ladder by 
consecutively adding blocks with 3 rungs (9 edges) to the ladder. 
From step 3 we now conclude that there are only two allowed configurations 
for the additional $S_1$-sites: they must be either both empty or
both occupied, since if just one of them is occupied the remaining 
configurations on the additional $S_2$-sites are exactly the ones of step 3. 
A simple computation shows that if both sites on $S_1$ are occupied, 
$H_{Q_2}$ has one non-trivial element with one dimer and if both sites 
on $S_1$ are empty, $H_{Q_2}$ has two non-trivial elements, both with 
two dimers. 

Now it is important to note that on the three additional rungs we find 
three non-trivial elements of the $H_{12}$, but two of them have $f_2=2$ 
and one has $f_2=1$. It can be shown that all three indeed belong to
$H_Q$ by going through the tic-tac-toe lemma step by step. This is a 
tedious computation, but it can be done.

Finally, we find the cohomology of the kagome ladder with open boundary 
conditions of length $n$, which corresponds to a ladder with $n+1$ rungs 
and $3n+1$ edges in total, by recursively adding rungs to the system. 
We thus obtain a recursion relation for the ground-state generating 
function $P_n(z) = \hbox{tr}_{\textrm{GS}}(z^F)$, which gives the Witten 
index for $z=-1$ and the total number of ground states for $z=1$:
\begin{equation}\nonumber
P_{n+3}(z) = 2 z^2 P_n(z) + z^3 P_{n-1}(z) ,
\end{equation}
with $P_0 = 0$, $P_1 = z$, $P_2 = 2z^2$, $P_3 = z^3$. 
Instead of drawing conclusion from here, let us
picture the above in terms of tiles. From step 4 we conclude that we 
can cover the ladder with three tiles, two of size 9 (i.e. 9 edges) 
containing 2 dimers and one of size 12 containing 3 dimers. From this picture
we obtain the same recursion relation provided that we allow four 
initial tiles corresponding to the initial conditions of the recursion 
relation above. Furthermore, we can see directly that the window of filling
fraction of the tiles runs from 2/9 to 1/4. Using the
recursion relation, we find that the ground-state entropy
is set by the largest solution $\lambda_{\textrm{max}}$ of the 
characteristic polynomial $\lambda^4 - 2\lambda - 1 = 0$,
giving $S_{\textrm{GS}}/L = (\ln \lambda_{\textrm{max}})/3 = 0.1110 \dots$.

\subsection{2D triangular lattice}\label{triangle}

The ground state structure of the supersymmetric model on the
2D triangular lattice is not fully understood. Nevertheless,
it is clear that ground states occur in a finite window of 
filling fractions $\nu=f/L$ and that there is extensive ground 
state entropy. These features seem to be generic for 2D
lattices, as they have been observed for many examples such
as hexagonal, kagome, etc. (2D square being an important 
exception) \cite{vE}.

\begin{figure}
\centering
\resizebox{0.5\columnwidth}{!}{%
\includegraphics{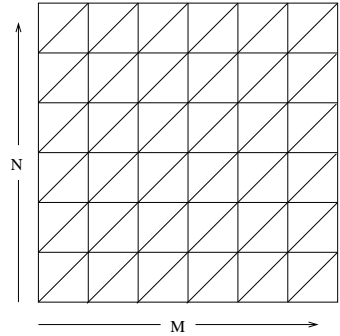}
}
\caption{The $M\times N$ triangular lattice has periodic boundary 
conditions along the directions of the two arrows.}
\label{triangularlattice}
\end{figure}

\begin{table}
\caption{Witten Index for the $M \times N$ triangular lattice.}
\centering
{\scriptsize
\begin{tabular}{r|rrrrrrrr}
   & 1 &   2 &   3 &     4 &      5 &      6 &       7  \\
\hline
 1 & 1 &   1 &   1 &     1 &      1 &      1 &       1  \\
 2 & 1 &  -3 &  -5 &     1 &     11 &      9 &     -13  \\
 3 & 1 &  -5 &  -2 &     7 &      1 &    -14 &       1  \\
 4 & 1 &   1 &   7 &   -23 &     11 &     25 &     -69  \\
 5 & 1 &  11 &   1 &    11 &     36 &    -49 &     211  \\
 6 & 1 &   9 & -14 &    25 &    -49 &   -102 &     -13  \\
 7 & 1 & -13 &   1 &   -69 &    211 &    -13 &    -797  \\
 8 & 1 & -31 &  31 &   193 &   -349 &   -415 &    3403  \\
 9 & 1 &  -5 &  -2 &   -29 &    881 &   1462 &   -7055  \\
10 & 1 &  57 & -65 &  -279 &  -1064 &  -4911 &    5237  \\
11 & 1 &  67 &   1 &   859 &   1651 &  12607 &   32418  \\
12 & 1 & -47 & 130 & -1295 &   -589 & -26006 & -152697  \\
13 & 1 &-181 &   1 &   -77 &  -1949 &  67523 &  330331  \\
14 & 1 & -87 &-257 &  3641 &  12611 &-139935 & -235717  \\
15 & 1 & 275 &  -2 & -8053 & -32664 & 272486 & -1184714 \\
\end{tabular}
}
\label{tab:triang}
\end{table}

In Tab.~\ref{tab:triang} we show the Witten indices for the 
$M \times N$ triangular lattice, with periodic boundary conditions 
applied along two axes of the lattice (see Fig.~\ref{triangularlattice}). 
The exponential growth of the index is 
clear from the table. To quantify the growth behavior, one may 
determine the largest eigenvalue $\lambda_N$ of the row-to-row
transfer matrix for the Witten index on size $M \times N$.
This gives
\begin{eqnarray}
&& |W_{M,N}| \sim (\lambda_N)^M + (\bar{\lambda}_N)^M \ , \quad
\lambda_N\sim \lambda^N 
\nonumber\\[2mm]
&& |\lambda|\sim 1.14 \ , \quad {\rm arg}(\lambda)\sim 0.18 (\pi)
\end{eqnarray}
leading to a ground state entropy per site of
\begin{equation}
\frac{S_{\rm GS}}{MN} \geq \frac{1}{MN} \log |W_{M,N}|
\sim \log |\lambda| \sim 0.13 \ .
\end{equation}
The argument of $\lambda$ indicates that the asymptotic
behavior of the index is dominated by configurations
with filling fraction around $\nu=0.18$.

In a most interesting mathematical analysis \cite{Jo}, Jonsson has 
shown that for a sufficiently large triangular lattice (with open BC) 
ground states occur for all rational numbers in the range
\begin{equation}
1/7 \leq \nu \leq 1/5 \ .
\end{equation}
His analysis is based on an effective geometric picture involving 
so-called cross-cycles, which however is less explicit than 
the one that has been worked out for the case of the 2D 
square lattice (see below). There is a clear challenge to develop 
this picture further to the point that the growth behavior of the
Witten index, and of the number of ground states, can be 
given in closed form.

\subsection{2D square lattice}\label{square}

\begin{table}
\caption{Witten Index for $M \times N$ square lattice.}
\centering
{\scriptsize
\begin{tabular}{r|rrrrrrrrrrrrr}
 & 1 & 2 & 3 & 4 & 5 & 6 & 7 & 8 & 9 & 10 & 11 & 12\\
\hline
1  & 1 &  1 & 1 & 1 &  1 &  1 & 1 & 1 & 1 & 1  & 1 & 1  \\
2  & 1 & -1 & 1 & 3 &  1 & -1 & 1 & 3 & 1 & -1 & 1 & 3  \\
3  & 1 &  1 & 4 & 1 &  1 &  4 & 1 & 1 & 4 & 1  & 1 & 4  \\
4  & 1 &  3 & 1 & 7 &  1 &  3 & 1 & 7 & 1 & 3  & 1 & 7  \\
5  & 1 &  1 & 1 & 1 & -9 &  1 & 1 & 1 & 1 & 11 & 1 & 1  \\
6  & 1 & -1 & 4 & 3 &  1 & 14 & 1 & 3 & 4 & -1 & 1 & 18 \\
7  & 1 &  1 & 1 & 1 &  1 &  1 & 1 & 1 & 1 & 1  & 1 & 1  \\
8  & 1 &  3 & 1 & 7 &  1 &  3 & 1 & 7 & 1 & 43 & 1 & 7  \\
9  & 1 &  1 & 4 & 1 &  1 &  4 & 1 & 1 & 40& 1  & 1 & 4  \\
10 & 1 & -1 & 1 & 3 & 11 & -1 & 1 & 43& 1 & 9  & 1 & 3  \\
11 & 1 & 1  & 1 & 1 & 1  & 1  & 1 & 1 & 1 & 1  & 1 & 1  \\
12 & 1 & 3  & 4 & 7 & 1  & 18 & 1 & 7 & 4 & 3  & 1 &166 \\
13 & 1 & 1  & 1 & 1 & 1  & 1  & 1 & 1 & 1 & 1  & 1 & 1  \\
14 & 1 & -1 & 1 & 3 & 1  & -1 &-27& 3 & 1 & 69 & 1 & 3  \\
15 & 1 & 1  & 4 & 1 & -9 & 4  & 1 & 1 & 4 & 11 & 1 & 4  \\
\end{tabular} 
}
\label{tab:square}
\end{table}

Numerical studies \cite{FSvE}
of the Witten index of the square lattice revealed 
a very different behavior (see Tab.~\ref{tab:square}). At first glance 
one notices that it does not grow exponentially with the system size.
In fact, more detailed investigation of these studies led to two 
conjectures \cite{FSvE} for which a proof was found by Jonsson 
\cite{Jo}. We state one of these results here: 
\begin{quote}
for an $M\times N$ square lattice with periodic boundary conditions 
in both directions, $W=1$ when $M$ and $N$ are coprime.
\end{quote}

\begin{figure}
\centering
\resizebox{0.8\columnwidth}{!}{%
\includegraphics{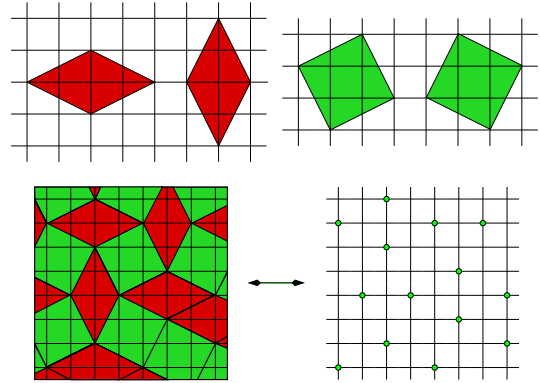}
}
\caption{Tilings of the 2D square lattice. Above: the four different
rhombi. Below: mapping between tiles
and hard-core fermions.}
\label{rhombi}
\end{figure}

Extending this work, Jonsson found a 
general expression for the Witten index $W_{u,v}$ of hard-core fermions 
on the square lattice with periodic boundary conditions given by
the vectors $u=(u_1,u_2)$ and $v=(v_1,v_2)$. 
The $M\times N$ square lattice is now a specific case with 
$u=(M,0)$ and $v=(0,N)$ 
(for an extension of this work to other families of grid graphs see 
\cite{BMLN}). A crucial step in \cite{Jo} is the introduction
of rhombus tilings of the square lattice. It is shown that the 
trace in the Witten index can be restricted to configurations that 
can be mapped to coverings of the plane with the four rhombi or 
tiles shown in Fig. \ref{rhombi}. 
Note that the sides of these rhombi, which connect the hard-core fermions, 
are in agreement with the heuristic 3-rule. Furthermore, two of the rhombi 
have area 4, whereas the other two have area 5. A covering with either of 
these rhombi alone thus corresponds to a filling fraction of 1/4 or 1/5,
respectively. 

To state Jonsson's results for the Witten index we introduce the following
notations. We denote by $R_{u,v}$ the family of tilings of the plane 
with boundary conditions given by $u=(u_1,u_2)$ and 
$v=(v_1,v_2)$. Furthermore $|R^+_{u,v}|$ and $|R^-_{u,v}|$ are 
the number of tilings of this plane with an even and an odd number 
of tiles, respectively. Finally, we define  
\begin{equation}
\theta_d \equiv \left\{ \begin{array}{ll}
2 & \textrm{if $d=3k$, with $k$ integer}\\
-1 & \textrm{otherwise.}
\end{array} \right.
\end{equation}
The expression for the Witten index then reads \cite{Jo}
\begin{equation}
W_{u,v}= - (-1)^d \theta_d \theta_{d^*} + |R^+_{u,v}| - |R^-_{u,v}|,
\end{equation}
where $d\equiv \hbox{gcd}(u_1-u_2,v_1-v_2)$ and 
$d^* \equiv \hbox{gcd}(u_1+u_2,v_1+v_2)$. 
It can be shown that the Witten index grows exponentially with the linear 
size (not the area) of the 2D lattice. Detailed results for the case of
diagonal boundary conditions have been given in \cite{Jo2}. Further studies of the Witten index
transfermatrix for the square lattice with diagonal and free boundary conditions by R.J.~Baxter \cite{RJB}
have led to an additional set of conjectures.

As we already mentioned in Sect. \ref{beyond1D}, the geometric picture 
in terms of tilings is useful beyond the computation of the Witten index. 
It also provides a way to determine a window of filling fractions in which
supersymmetric ground states can be found. The result is \cite{Jo} that 
for large enough square lattices (with open BC) ground states exist for all 
rational fillings in the range
\begin{equation}
1/5 \leq \nu \leq 1/4 \ .
\end{equation}

While it is clear that the effective geometric picture in terms
of rhombus tilings goes a long way characterizing the supersymmetric
ground states, it has until now failed to give complete results 
for the ground state partition sum for the supersymmetric model on the 2D
square lattice. For this issue, and for many others, it is important 
that the results presented in this section hold for the square lattice with 
any kind of periodic boundary conditions. This also includes semi-2D 
lattices, i.e. various ladders and even the 1D chain. These lattices are 
a good arena to further investigate properties of the supersymmetric 
fermion models, both analytically and numerically \cite{FHHS}.  

\section{Conclusion}
\label{conclusion}

The analysis of strongly correlated fermions on lattices in dimension
$D>1$ is a notoriously difficult problem, for which very few exact
results have been obtained. At the same time, the problem is highly 
relevant, as it holds the key to the behavior of correlated electrons 
in quasi-2D materials. We have here presented various exact results 
for the ground state structure of a fermion lattice model with an exact 
supersymmetry. In particular, we have demonstrated the remarkable feature 
of superfrustration, which this model possesses on generic 2D lattices.

In our discussion of the various examples of superfrustration, we 
mostly focused on specifying the number of supersymmetric 
ground states and the fermion number (or filling fraction) where they 
occur. Clearly, one would like to understand better various properties 
of these states, as well as the quantum phases they give rise to
when parameters are perturbed away from the supersymmetric point. 

The supersymmetric ground states on the 1D chain are quantum critical and
as such described by a superconformal field theory. For a more general
class of supersymmetric 1D models \cite{FNS} (where the nearest neighbor 
exclusion rule is softened) the situation is akin to that of higher-$S$ 
spin chains: the models are gapped but go critical if interaction 
parameters are tuned to specific values. For the 2D models presented here, 
the issue of quantum criticality is under investigation \cite{FHHS}. 
While supersymmetry alone certainly does not imply quantum 
criticality, it is clear that the balancing between kinetic and 
interaction terms that is implied by supersymmetry steers one
into regions of parameter space where charge order and Fermi liquid 
behavior compete.

\vskip 5mm

\noindent
{\it Acknowledgements.}
We thank Paul Fendley and Hendrik van Eerten for collaboration on the 
research that is here reviewed. We acknowledge financial support through
a PIONIER grant of NWO of the Netherlands and through the Research 
Networking Programme INSTANS of the ESF.

%
%

\end{document}